\documentclass{tMOP2e}
\usepackage{epsfig}
\usepackage{amsmath}
\usepackage{mathrsfs}
\usepackage{textcmds}
\newcommand{\be}{\begin{equation}}
\newcommand{\ee}{\end{equation}}
\newcommand{\bea}{\begin{eqnarray}}
\newcommand{\eea}{\end{eqnarray}}
\newcommand{\ket}[1]{\left|#1\right\rangle}
\newcommand{\bra}[1]{\left\langle #1\right|}

\newcommand{\bc}{\begin{center}}
\newcommand{\ec}{\end{center}}

\newcommand{\av}[1]{\left\langle #1\right\rangle}

\renewcommand{\(}{\left(}
\renewcommand{\)}{\right)}

\newcommand{\forget}[1]{}
\newcommand{\re}{{\rm e}}
\newcommand{\ri}{{\rm i}}

\begin{document}
\doi{10.1080/09500340xxxxxxxxxxxx}
 \issn{1362-3044}
\issnp{0950-0340}
\jyear{2006}
\markboth{Polarization preserving QND photodetector}{Polarization preserving QND photodetector}
\title{Polarization preserving quantum nondemolition photodetector}
\author{K. T. Kapale\\ Jet Propulsion Laboratory,
California Institute of Technology, Mail Stop 126-347, 4800 Oak Grove Drive, 
Pasadena, California 91109-8099}
\received{Jan 28, 2006}
\maketitle
\begin{abstract}
A polarization preserving quantum nondemolition photodetector is proposed based on nonlinearities obtainable through quantum coherence effects. An atomic level scheme is devised such that in the presence of strong linearly polarized drive field a coherent weak probe field acquires a phase proportional to the number of photons in the signal mode immaterial of its polarization state. It is also shown that the unavoidable phase-kicks resulting due to the measurement process are insensitive to the polarization state of the incoming signal photon. It is envisioned that such a device would have tremendous applicability in photonic quantum information proposals where quantum information in the polarization qubit is to be protected.
\end{abstract}

\section{Introduction}
Quantum coherence effects such as coherent population trapping (CPT)~\cite{Arimondo:1996} can give rise to strong higher order optical nonlinearities~\cite{Zubairy:2002}. In this context it can be noted that the optical nonlinearities of Kerr-type have been shown to be useful for quantum nondemolition (QND) measurement of the photon number~\cite{Imoto:1985}. Owing to recent research interest in the area of quantum information processing with photons as qubits,  quantum nondemolition detectors have found several applications such as quantum memory~\cite{Gingrich:2003}, quantum repeaters~\cite{Jacobs:2002} and preservation of photonic quantum information through quantum zeno effect~\cite{Spedalieri:2005}. In this context the quantum coherence effect based QND device was studied thoroughly to show that single photon sensitivities are achievable with atomic gases as active media~\cite{Munro:2005}. It can, nevertheless, be realized that atomic systems as active media for obtaining the Kerr type optical nonlinearities have their own peculiarities. The schemes invariably turn out to be sensitive the polarization state of the signal photons, thus affecting the quantum information carried by them.
This disadvantage is specially important in the context of quantum communication where quantum information is encoded in the polarization degrees of freedom. Thus, for maintaining the polarization information the QND scheme should not be dependent on the polarization of the photon that is being measured and this polarization state should not change.  Munro et al.~\cite{Munro:2005} have proposed a simplistic scheme for achieving the polarization independence for the QND scheme through usage of two different QND devices for two perpendicular polarization modes. This scheme has problems of its own as discussed later.  Polarization insensitive QND operation has also shown to be possible through nonlinearities induced by projective measurements within the realm of linear optical manipulations of the signal photon~\cite{Kok:2002}. The linear optical techniques, however, are probabilistic. Thus, repeated applications of such QND devices would ultimately reduce the success probability of the protocol tremendously. In this context, the approach proposed here would give highly efficient  and polarization preserving QND device.

The article is organized as follows. In sec 2 a brief review of how Kerr type nonlinearities can be achieved through atomic coherence effects  and how this nonlinearity is useful for devising a QND photodetector is given.  In  sec 3
the model is discussed in detail and derivations of the effective hamiltonian is provided. In the following section how the effective hamiltonian allows polarization preserving QND operation is discussed along with a simple noise analysis showing that the phase kicks arising due to the detection process do not affect the polarization information of the photonic qubit. Finally, conclusions are provided.

\section{Kerr nonlinearity through quantum coherence effects}
In this section we briefly review how the Cross-Kerr nonlinearity can be obtained through  quantum coherence effects and how to employ this nonlinearity to achieve  a QND measurement of photon number of a signal light.
For simplicity it is assumed that the systems operate in the cw regime and no attention is paid to special requirements of pulse-propagation through atomic media. Such considerations exist in the literature and could be applied to the problem at hand in a straightforward manner.

\begin{figure}[ht]
\centerline{\includegraphics[scale=0.4]{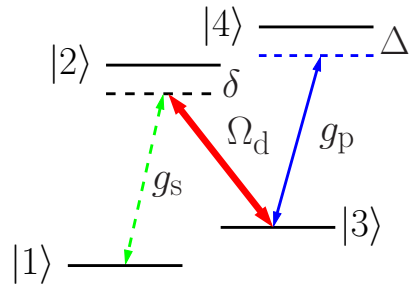}}
\caption{\label{Fig:N}$N$-type levelscheme for generation of cross-Kerr nonlinearity}
\end{figure}

We consider a four level $N$-type atomic level scheme as shown in Fig.\ref{Fig:N}.  
Following on the lines of Ref.~\cite{Zubairy:2002}
it can be shown that for an atom initially in the state $\ket{1}$ the effective Hamiltonian could be written as
\begin{equation}
{\mathscr H}_{\rm eff}= - \hbar \frac{\xi_{\rm s}^2 \hat{a}_{\rm s}^\dagger\,\hat{a}_{\rm s}}{\Delta}\frac{\xi_{\rm p}^2 \hat{a}_{\rm p}^\dagger\,\hat{a}_{\rm p}}{|\Omega_{\rm d}|^2} \sum_{n_s, n_p}^{\infty}\ket{1,n_{\rm s},n_{\rm p}}\bra{1,n_{\rm s},n_{\rm p}}\,.
\label{Eq:HeffN}
\end{equation}
Here $\Omega_{\rm d}$ is the Rabi frequency of the strong drive field and the signal and probe fields are taken to be weak quantum fields 
with the Rabi frequencies defined through
\begin{equation}
\label{Eq:SignalProbeRabi}
\hat{g}_i = \sqrt{\frac{2 \pi \wp_{i}^2 \nu_{i}}{\hbar V_{i}}}\, \hat{a}_{i}=\xi_{i}\hat{a}_{i}  \qquad i = {\rm s},{\rm p}\,,
\end{equation}
where $\wp_{\rm s, p}$ are the dipole moments of the signal ($\ket{1}$--$\ket{2}$) and probe ($\ket{3}$--$\ket{4}$) transitions respectively, $\nu_{\rm s,p}$ are the corresponding field frequencies and $V_{\rm s, p}$ correspond to the quantization volume of the modes, and $\hat{a}_{\rm s, p}$ and $\hat{a}^\dagger_{\rm s, p}$ are the annihilation and creation operators. It can be noted that both the signal and probe fields are taken to be quantum fields that are eigenstates of the number operators $\hat{n}_{\rm s}=\hat{a}_{\rm s}^\dagger\,\hat{a}_{\rm s}$ and $\hat{n}_{\rm p}=\hat{a}_{\rm p}^\dagger\,\hat{a}_{\rm p}$ respectively.

 As shown in the figure the signal and drive fields are in two photon resonance with single photon detuning given by $\delta$ while the probe field is detuned by $\Delta$ from the  probe transition. The detuning $\delta$ could very well be zero in the case discussed in this section; however, it is crucial for the polarization preserving scheme. The parameters are chosen such that $\Delta, \delta\gg |\Omega_{\rm d}| \gg |g_{i}|$. Within this choice of the atom-field interaction parameters the usual CPT dark state $(\xi_{\rm s}\sqrt{n_{\rm s}} \ket{3, n_{\rm s}-1}- \Omega_{\rm d} \ket{1,n_{\rm s}}) /\sqrt{\xi_{\rm s}^2 n_{\rm s} + |\Omega_{\rm d}|^2}$ for the three level system formed by levels $\ket{1}$, $\ket{2}$, and $\ket{3}$ is perturbed  in the presence of the highly detuned probe field. Strictly speaking there is not dark state after the addition of the probe field; however $\ket{1, n_{\rm s}, n_{\rm p}}$ is a quasidark state corresponding to the eigenvalue that goes to zero when the probe detuning, $\Delta$, is infinite. This eigenvalue is the term appearing just before the summation sign in Eq.~\eqref{Eq:HeffN}~\footnote{Interested reader can browse through the appendix of Ref.~\cite{Zubairy:2002} for the derivation of this result.}.

\begin{figure}[ht]
\centerline{\includegraphics[scale=0.5]{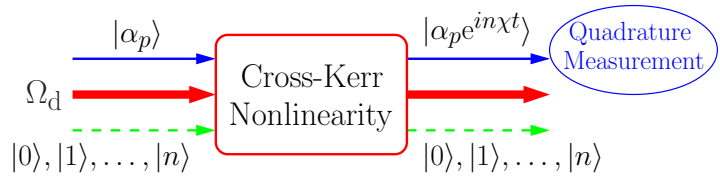}}
\caption{\label{Fig:Cross-Kerr}Cross-Kerr Nonlinearity for QND measurement}
\end{figure}

It is clear that the Hamiltonian of Eq.~\eqref{Eq:HeffN} gives cross-Kerr nonlinearity for the probe field based on the presence of the signal photon in the presence of strong drive field (with initial atomic state $\ket{1}$ that is an eigenstate), so that it can be written as the QND Hamiltonian~\cite{Imoto:1985}
\begin{equation}
\mathscr{H}_{\rm QND} = \hbar \chi \hat{a}_{\rm s}^\dagger\,\hat{a}_{\rm s}
\hat{a}_{\rm p}^\dagger\,\hat{a}_{\rm p}
\end{equation}
with 
\begin{equation}
\chi = - \frac{\xi_{\rm s}^2 \xi_{\rm p}^2 }{\Delta |\Omega_{\rm d}|^2}\,.
\end{equation}
Thus, if the signal field contains $n_{\rm s}$ photons and the probe field is in a coherent state with amplitude $\ket{\alpha_{\rm p}}$ initially, the cross-Kerr optical nonlinearity causes the combined system to evolve into
\begin{equation}
\ket{\Psi(t)}_{\rm out} = \re^{\ri \chi \hat{a}_{\rm s}^\dagger\,\hat{a}_{\rm s}
\hat{a}_{\rm p}^\dagger\,\hat{a}_{\rm p}} \ket{n_{\rm s}}\ket{\alpha_{\rm p}} = \ket{n_{\rm s}}\ket{\alpha_{\rm p}\re^{\ri n_{\rm s} \chi t}\,.
}
\end{equation}
Thus, the signal fock state is unaffected by the interaction and the coherent state picks up a phase shift directly proportional to the number of photons $n_{\rm s}$ in the signal mode. This phase shift can be measured using a homodyne measurement as depicted schematically in Fig.~\ref{Fig:Cross-Kerr}, so the number of photons in the signal mode can be inferred directly.

It should, however, be noted that real experimental implementation of such a scheme would use atomic transitions as shown in the Fig.\ref{Fig:N} that are always polarization dependent. Thus, employing this QND scheme for quantum information applications with polarization encoded photonic qubits,  would most certainly destroy the qubit. The question then arises---how to perform a QND measurement of the photon without affecting its polarization state. Munro {\it et al.}~\cite{Munro:2005} have studied this photon number resolving QND photodector in great details and they have shown it to be highly efficient and accurate. 

To achieve polarization independence Munro {\it et al.} propose to split the signal beam through a polarizing beam splitter in two components and to use two independent but identical cross-Kerr nonlinearities to change the phase of the coherent probe field in a sequential manner and then to combine the signal polarization components together to achieve polarization independent QND device. It can be quickly seen that in practice it would be extremely difficult to satisfy the requirements of this setup. To illustrate, such a scheme would need exactly identical Kerr media giving identical nonlinearities for two perpendicular modes of signal-field polarizations. Given that alkali atomic gases are ideally suited for implementing the cross-Kerr nonlinearities, it would be almost impossible to match two atomic cells to have exactly identical properties. Moreover, the scheme might only work for a chosen basis of polarization dictated by the properties of the polarizing beam splitters. Quantum communication proposals normally require change of polarization basis of the signal photon from time to time to ensure security, in such an event an alternative approach is needed for polarization preserving QND device.

In the next section such a scheme is proposed and it is shown that it gives rise to a new type of a QND Hamiltonian.

\section{Polarization preserving QND scheme}

\begin{figure}[ht]
\centerline{\includegraphics[scale=0.5]{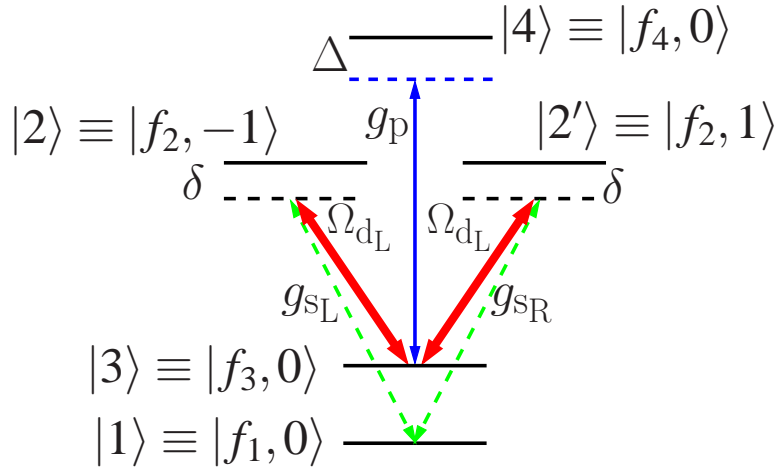}}
\caption{\label{Fig:PP}  Level scheme for achieving polarization preserving Cross-Kerr nonlinearity}
\end{figure}

The level scheme for achieving polarization independence
is depicted in the Fig.~\ref{Fig:PP}. The transitions are chosen such that they are excited by definite polarization components of the coupling fields. The atomic levels are identified by their specific hyperfine quantum numbers $f$ and $m_f$; as shown only definite values of $m_f$ are necessary and $f$ can be chosen with some freedom. The drive and signal transitions are now two-fold and contain either left or right circularly polarized (LCP or RCP) modes. The requirements are such that both the polarization components should have identical Rabi-frequency amplitudes. Namely, $|\Omega_{\rm d_L}|=|\Omega_{\rm d_R}|=\Omega_{\rm d}$ and $|g_{\rm d_L}|=|g_{\rm d_R}|=g_{\rm s}$. This can be achieved by choosing a linearly polarized drive field. As  linear polarization is just an equal-weight superposition of LCP and RCP polarizations, single drive field is sufficient to excite both the $\ket{3}$--$\ket{2}$ and  $\ket{3}$--$\ket{2'}$ transitions. For simplicity, the $\ket{3}$--$\ket{4}$ transition is taken to be coupled by linearly polarized probe field thus it requires $f_4=0$ so that it does not give rise to two opposite circularly polarized transitions. In reality, if the probe field instead couples to two levels through LCP and RCP transitions, same as the drive field, the whole scheme would still work as long as the two corresponding Rabi-frequencies have the same amplitude. Thus, there is sufficient freedom in the choice of atomic species for employing this scheme.
It can be easily verified that such a scheme can be easily obtained for atomic Cs.

The Hamiltonian of the system is given by
\begin{align}
\mathscr{H}&= \hbar \Delta \ket{4}\bra{4} + \hbar \delta \(\ket{2}\bra{2}+\ket{2'}\bra{2'}\)+ 
\left\{ 
\hbar \hat{g}_{\rm s_{\rm R}} \ket{2'}\bra{1}
+ \hbar \hat{g}_{\rm s_{\rm L}} \ket{2}\bra{1}
\right.\\ 
& + \left.\hbar \Omega_{\rm d_{L}}\ket{2}\bra{3} 
+\hbar \Omega_{\rm d_{R}} \ket{2'}\bra{b_2}
+\hbar \hat{g}_{\rm p} \ket{4}\bra{3}  + {\rm H. c.}\right\}\,.
\end{align}
The drive, signal and probe Rabi frequencies are defined for their respective polarization modes as:
\begin{align}
\hat{g}_{\rm s_{L}} = \xi_{\rm s_{L}} \hat{a}_{\rm s_{L}}, \hat{g}_{\rm s_{R}} = \xi_{\rm s_{R}} \hat{a}_{\rm s_{R}}, \hat{g}_{\rm p} = \xi_{\rm p} \hat{a}_{\rm p}, \hbar \Omega_{\rm d_{L}} = \vec{\wp}_{23}\cdot\vec{E}_{\rm d}, \hbar \Omega_{\rm d_{R}} = \vec{\wp}_{2'3}\cdot\vec{E}_{\rm d}
\end{align}
with $\vec{\wp}_{ij}$ is the dipole moment of the atomic transition between levels $i$ and $j$. More clearly, $\vec{\wp}_{23}=\wp_{23}(\hat{x} + \ri \hat{y})$ and $\vec{\wp}_{2'3}=\wp_{2'3}(\hat{x} - \ri \hat{y})$. Thus, if  the drive electric field is taken to be $\vec{E}_{\rm d} = E_{\rm d} \hat{x}$ it couples both the LCP and RCP transitions equally. The atomic transitions are chosen such that $|\Omega_{\rm d_{L}}|=|\Omega_{\rm d_{R}}|=\Omega_{\rm d}$
 and $|\xi_{\rm s_{L}}| = |\xi_{\rm s_{R}}| = \xi_{\rm s}$ which is a common occurrence in atomic transitions between different hyperfine manifolds.
The detunings are defined as $\Delta = \omega_{43} - \nu_{\rm p}$  and $\delta = \omega_{2'1}-\nu_{\rm s} = \omega_{21} - \nu_{\rm s}= \omega_{2'3} - \nu_{\rm d} = \omega_{23} - \nu_{\rm d}$ with $\omega_{ij}$ being the atomic energy difference between levels $\ket{i}$ and $\ket{j}$ and $\nu_{\rm s, p, d}$ are the field frequencies on the signal, probe and drive transitions respectively. It is imperative to point out that the signal and drive field in either polarization excite a two-photon transition between states $\ket{1}$ and $\ket{3}$, but due to the presence of large $\delta$ they not give rise to a single photon transitions. This is especially important if the incoming signal photon is say left-circularly polarized; then only the $N$-type scheme involving levels $\{\ket{1},\ket{2},\ket{3},\ket{4}\}$ should take part in the dynamics to give an appropriate QND scheme. For the right-circularly polarized signal field, on the other hand, the $N$-type scheme formed by $\{\ket{1},\ket{2},\ket{3},\ket{4}\}$ gives the required QND scheme. These individual QND schemes work exactly as discussed in Sec 2.  It is important to see how the system would behave for a linearly polarized signal field that is a superposition of two circularly polarized components.

To clarify the kind of approximations involved the evaluation of the effective Hamiltonian is considered in little more details here.
The basis set used for matrix representation of the Hamiltonian is $\{ \ket{1,n_{\rm s}, n_{\rm p}}, \ket{2,n_{\rm s_{\rm L}}-1,n_{\rm p}}, \ket{2',n_{\rm s_{\rm R}}-1,n_{\rm p}}, \ket{3,n_{\rm s}-1,n_{\rm p}}, \ket{4, n_{\rm s}, n_{\rm p}}\}$, with $n_{\rm s}=n_{\rm s_{\rm L}}+n_{\rm s_{\rm R}}$. It should be noted that the state $\ket{3,n_{\rm s}-1,n_{\rm p}}$ does not specify which route, absorption of LCP or RCP polarized signal photon, brought the atom to it from the initial ground state $\ket{1,n_{\rm s}, n_{\rm p}}$. This, feature is essentially the one that gives effective polarization preserving capability of the scheme as it will become clear later. 

The aim is to find the perturbed dark state of the Hamiltonian such that in the limit  $\Delta\rightarrow \infty$ the eigenvalue  goes to zero. Thus, in essence we are looking for the smallest eigenvalue of the Hamiltonian. The 
secular equation can be written as
\begin{equation}
-\lambda^5 + a \lambda^4 + b \lambda^3 + c \lambda^2 + d \lambda + e = 0\,,
\end{equation} 
with
\begin{align}
a & = (2 \delta + \Delta)\,, \nonumber \\
b & = - \delta^2 - 2 \delta \Delta + \xi_{\rm s}^2(\hat{n}_{\rm s_{\rm L}}+\hat{n}_{\rm s_{\rm R}}) + 2 |\Omega_{\rm d}|^2 +  \xi_{\rm p}^2 n_{\rm p}\,, \nonumber \\
c & = - (\delta+\Delta)  \xi_{\rm s}^2 (\hat{n}_{\rm s_{\rm L}}+\hat{n}_{s_{\rm R}})  + \delta^2 \Delta - 2 (\delta+\Delta) |\Omega_{\rm d}|^2  - 2 \delta \,\xi_{\rm p}^2 n_{\rm p}\,, \nonumber \\
d &=\delta \Delta(  \xi_{\rm s}^2(\hat{n}_{\rm s_{\rm L}}+\hat{n}_{\rm s_{\rm R}}) + 2 |\Omega_{\rm d}|^2 )- \xi_{\rm s}^2(\hat{n}_{\rm s_{\rm L}}+\hat{n}_{\rm s_{\rm R}})\, \xi_{\rm p}^2 n_{\rm p} + \delta^2  \xi_{\rm p}^2 n_{\rm p}\,,\nonumber \\
e&=\delta\,  \xi_{\rm s}^2(\hat{n}_{\rm s_{\rm L}}+\hat{n}_{\rm s_{\rm R}})\,  \xi_{\rm p}^2 n_{\rm p}\,.
\end{align}
Here,  the polarization dependent number operators $\hat{n}_{\rm s_{\rm L}}$ and $\hat{n}_{\rm s_{\rm R}}$ are left in the operator format as the basis states do not specify the polarization state of the incoming photon. It is easy to see that no matter what the polarization state of the signal photon $\hat{n}_{\rm s_{\rm L}} + \hat{n}_{\rm s_{\rm R}}= n_{\rm s}$. 

Thus, the two number operators coming in this linear combination form gives this Hamiltonian its polarization insensitive property. 

It is also easy to see that $\ket{n_{\rm s}}$ is an eigenstate of the operator $\hat{n}_{\rm s_{\rm L}} + \hat{n}_{\rm s_{\rm R}}$. 

Approximate solutions for the quintic equation can be obtained within the approximation $\Delta,\delta\gg|\Omega_{\rm d}|\gg\xi_{\rm p}\gg \xi_{\rm s}$. Thus, the largest eigenvalue is the nonzero solution of $-\lambda^5 + a\lambda^4=0$, i.e., $\lambda_{l}\approx2 \delta + \Delta$ with the eigenstate approximately given by $\ket{a_2, n_s-1,n_p-1}$. The next set of three eigenvalues could be obtained by solving $a \lambda^4 + b\lambda^3 + c\lambda^2+ d\lambda=0$. The smallest eigenvalue can be quickly shown to be given by
\begin{equation}
\lambda_s \simeq -\frac{e}{d}\simeq
-\frac{ \xi_{\rm s}^2(n_{s_{\rm L}}+n_{s_{\rm R}})\, \xi_{\rm p}^2 n_{\rm p} }{\Delta |\Omega_{\rm d}|^2}
\forget{=
-\frac{(\hat{\alpha}_{s_{\rm L}}^\dagger \hat{\alpha}_{s_{\rm L}} +\hat{\alpha}_{s_{\rm R}}^\dagger \hat{\alpha}_{s_{\rm R}})\hat{\alpha}_p^\dagger \hat{\alpha}_p}{\Delta |\Omega_{\rm D}|^2}}
\end{equation}
and the corresponding eigenstate is $\ket{1,n_{\rm s},n_{\rm p}}$.
\forget{Thus noting that the initial state of the atom fields is of the form
$
\sum_{n_p}\ket{b_1,n_s,n_p}
$
the effective Hamiltonian can be written in terms of the lowest eigenstate as
\begin{equation}
\mathscr{H}_{\rm eff} = -\frac{\xi_{\rm s}^2\xi_{\rm p}^2(\hat{a}_{s_{\rm L}}^\dagger \hat{a}_{s_{\rm L}} +\hat{a}_{s_{\rm R}}^\dagger \hat{a}_{s_{\rm R}})\hat{a}_{\rm p}^\dagger \hat{a}_{\rm p}}{\Delta |\Omega_{\rm d}|^2} \sum_{n_p}\ket{1,n_{\rm s},n_{\rm p}}\bra{1,n_{\rm s},n_{\rm p}}
\end{equation}
}
Noting that the atom does not change its state and remains in $\ket{1}$ the effective Hamiltonian  can be written as 
\begin{equation}
\mathscr{H}_{\rm eff} = -\hbar\frac{\xi_{\rm s}^2\xi_{\rm p}^2(\hat{a}_{s_{\rm L}}^\dagger \hat{a}_{s_{\rm L}} +\hat{a}_{s_{\rm R}}^\dagger \hat{a}_{s_{\rm R}})\hat{a}_{\rm p}^\dagger \hat{a}_{\rm p}}{\Delta |\Omega_{\rm d}|^2}\,.
\label{Eq:HeffPP}
\end{equation}
In the next section some properties of this hamiltonian are analyzed and it is  shown how it gives polarization preserving QND operation.

\section{Analysis}
The primary requirements for a polarization preserving QND device are---(i) the nonlinearity should be independent of the polarization state of the incoming photon, (ii) the interaction should not change the polarization state of the photon and (iii) the imperative phase kicks arising due to the QND measurement be immaterial of the polarization state of the signal photon. 
In this section it is shown that the above requirements are indeed satisfied  by the scheme discussed in the earlier section.

The effective Hamiltonian of Eq.~\eqref{Eq:HeffPP} can be rewritten as
\begin{equation}
\mathscr{H}_{\rm PPQND} = \hbar \chi \({\hat{a}_{s_{\rm L}}}^\dagger \hat{a}_{s_{\rm L}} +\hat{a}_{s_{\rm R}}^\dagger \hat{a}_{s_{\rm R}}\) {\hat{a}_{\rm p}}^\dagger \hat{a}_{\rm p}
\label{Eq:HeffPP1}
\end{equation}
It can be easily seen that by changing the polarization basis from circular to linear polarization through the relations
\begin{equation}
\hat{a}_{\rm s_{L}} = \(\hat{a}_{\rm s_{H}} + \ri\, \hat{a}_{\rm s_{V}}\)/\sqrt{2}, \quad {\rm  and }\quad  \hat{a}_{\rm s_{R}} = \(\hat{a}_{\rm s_{H}} - \ri\, \hat{a}_{\rm s_{V}}\)/\sqrt{2}
\label{Eq:basis}
\end{equation}
the effective Hamiltonian can also be represented as
\begin{equation}
\mathscr{H}_{\rm PPQND} = \hbar \chi \({\hat{a}_{s_{\rm H}}}^\dagger \hat{a}_{s_{\rm H}} +\hat{a}_{s_{\rm V}}^\dagger \hat{a}_{s_{\rm V}}\) {\hat{a}_{\rm p}}^\dagger \hat{a}_{\rm p}
\label{Eq:HeffPP2}
\end{equation}
In fact, it can be verified  that the form of the Hamiltonian remains invariant under any unitary transformation applied to the polarization degrees of freedom. The implication of the equivalence of the two forms of the Hamiltonian in Eqs.~\eqref{Eq:HeffPP1} and~\eqref{Eq:HeffPP2}, that the QND operation is polarization independent and polarization preserving. Alternative way of understanding the polarization independence could be through the fact that single photon state of any arbitrary polarization is an eigenstate of the operator ${\hat{a}_{s_{\rm L}}}^\dagger \hat{a}_{s_{\rm L}} +\hat{a}_{s_{\rm R}}^\dagger \hat{a}_{s_{\rm R}}$, i.e., $\hat{n}_{\rm s_{L}} + \hat{n}_{\rm s_{R}}$ with unity as eigenvalue.

To show that the phase kicks imposed by the the QND measurement the method of Imoto et al~\cite{Imoto:1985} can be followed to show that
\begin{equation}
\av{(\Delta n_{{\rm s}_{i}})^2}_{\rm measured} \av{(\Delta \phi_{{\rm s}_{i}})^2}_{\rm imposed} =  1/4\,.
\end{equation}
where $i$ are the two mutually orthogonal polarization components.
 Thus, through the equivalence of Eqs.~\eqref{Eq:HeffPP1} and~\eqref{Eq:HeffPP2} it is clear that whatever the polarization of the incoming signal photon the phase kick received by the two mutually orthogonal basis components is the same  because
\begin{equation}
\av{(\Delta n_{{\rm s}_{i}})^2}_{\rm measured} = \av{(\Delta n_{{\rm s}_{j}})^2}_{\rm measured}\,
\end{equation}
where $\{i, j\}$ are the mutually orthogonal basis elements chosen to represent the polarization of the incoming photon.
Thus, the QND measurement process imparts an immaterial overall phase factor to the state of the signal photon. 

To illustrate, from the Fig.~\ref{Fig:PP}, a single photon which is either LC polarized or RC polarized would have the same phase kick as the atomic properties are chosen such that $|\Omega_{\rm d_L}|= |\Omega_{\rm d_R}|$ and $|g_{\rm s_L}| = |g_{\rm s_R}|$. It remains to be seen what the phase kick would be for a photon which is polarized in say a H-V polarization basis perpendicular basis. Through equation~\eqref{Eq:basis} it is clear that an horizontally or vertically  polarized photon could be expressed as a linear combination of two circularly polarized components. Since there is  only a single signal photon arriving at the  location of the atom, the atom with equal probability could see it as either LCP or RCP photon, either way the kick it receives is the same. Thus, the back-action of the  QND measurement on the phase of the photon is immaterial of its polarization state and it only gives an overall phase factor to the photonic state.  

\section{Conclusions}

It is shown that through a proper choice of atomic levelscheme a polarization preserving nonlinearity could be induced in an atomic medium based on the quantum coherence effects. The induced nonlinearity can be used to obtain a single photon sensitive QND photodetector that preserves the polarization state of the photon. A simple application of such a device could be to preserve polarization encoded quantum information in linear optical quantum information processors and quantum cryptographic systems. The author hopes that the study presented here would give rise to an interest in finding further applications of such a device.

\bibliographystyle{prsty}
\bibliography{QND}

\end{document}